# Valuation of Currency Options in Markets with a Crunch[1]

Abdulnasser Hatemi-J[i] and Youssef El-Khatib[ii]

i. Department of Finance, UAE University, P.O. Box 15551, Al-Ain, UAE.
   Email: *AHatemi@uaeu.ac.ae*
i. Department of Mathematical Sciences, UAE University, P.O. Box 15551, Al-Ain, UAE.
   Email: *Youssef_Elkhatib@uaeu.ac.ae*

Abstract

This work studies the valuation of currency options in markets suffering from a financial crisis. We consider a European option where the underlying asset is a foreign currency. We assume that the value of the underlying asset is a stochastic process that follows a modified Black-Scholes model with an augmented stochastic volatility. Under these settings, we provide a closed form solution for the option-pricing problem on foreign currency for the European call and put options. A mathematical proof is provided for the underlying solution. In addition, simulation results and an application are provided.

[1] This research is partially funded by a research grant provided by the UAE Univesity (i.e. Grant Nr. 31B028), which is gratefully appriciated.

## 1. Introduction

The evaluation of options is an integral part of modern financial risk management. Due to increasingly globalized financial markets currency options are increasingly used by multinational corporations and other institutions as well as individuals to neutralize the underlying exchange rate risk. It is widely agreed in the recent literature that the volatility of many exchange rates are increasing across time. This seems to be the case particularly during the last two decades. In the existing literature there is an option pricing formula for currency evaluation that is introduced by Biger and Hull (1983) as well as Garman and Kohlhagen (1983), which is based on the seminal work of Black and Scholes (1973). The goal of the current paper is to extend the existing formula for tackling the valuation of currency options in markets that are suffering from a potential financial crisis. We concentrate on a European option where the underlying asset is a foreign currency. We assume that the value of the underlying asset is a stochastic process that follows a modified Black-Scholes model with an augmented stochastic volatility. Under these settings, we derive formulas for the option-pricing problem on foreign currency for both European call and put options. The well-known Black and Scholes (1973) option pricing formula is routinely used for the valuation of different options. It is a well-established fact in the literature that the volatility of financial assets tend to increase during a financial crisis period. Due to the increasingly dominant globalization effect of the financial markets, the likelihood of spillover effects and the resulting contagion is higher than ever. As a consequence, the Black and Scholes (1973) formula might not perform accurately during a financial crisis. Thus, our suggested approach might be useful under these circumstances.

After this introduction the rest of the paper is organized as follows. In Section 2 the model for generating the price of the underlying currency, which also takes into account the effect of a potential crisis, is presented. Section 3 introduces the formula for valuation currency options along with the proof. Section 4 illustrates graphically the predicted trajectory for the foreign currency spot rate based on the standard approach as well as based on the suggested approach provided by the current paper. Section 5 provides an application. The last section offers the concluding remarks.

## 2. The Model

Following Black and Scholes (1973) we make the subsequent assumptions in order to derive the currency option pricing formula.

1. The short-term risk free rates for both domestic and the foreign markets, i.e. $r_d$ and $r_f$, are assumed to be known and constant.[2]

2. It is assumed that the underlying distribution of currency prices within any finite interval is lognormal.

3. No dividend payouts take place during the life time of the underlying option[3].

4. There are no transaction costs.

5. Short selling is possible.

Nevertheless, unlike the original Black and Scholes approach the variance of the original asset does not need to be constant and it can increase across time.

The following denotations are used in the current paper. Let the probability space be defined by $(\Omega, F, P)$. Let also that $(W_t)_{t \in [0,T]}$ represent a Brownian motion process with $(F_t)_{t \in [0,T]}$ being the natural filtration generated by $(W_t)_{t \in [0,T]}$. In addition, let $P$ signify the risk-neutral probability. In this case the data generating process for the foreign currency at time $t$ that is determined via $P$, defined as $S_t$, is the following stochastic differential equation:

$$dS_t = (r_d - r_f) S_t dt + \left(\sigma S_t + \beta e^{(r_d - r_f)t}\right) dW_t, \tag{1}$$

---

[2] For currency option pricing with stochastic interest rate see Amin and Jarrow (1991). See also El-Khatib and Hatemi-J (2012, 2013) and El-Khatib, Hajji, Al-Refai (2013) for price sensitivities calculation during crisis as well as for processes with jumps.
[3] This assumption can be relaxed if the underlying market price of the asset is discounted continuously with regard to the dividends.

where $t \in [0,T]$, $S_0 > 0$ and $\beta$ is a constant. The denotation $\sigma$ signifies the volatility of the original asset. Note that $r_d$ is the domestic interest rate and $r_f$ is the foreign interest rate. Equation (1) accounts of the post-crash (crisis) effect via the parameter $\beta > 0$.

**Proposition 1** *For $0 \leq t \leq T$, the equation (1) has the following solution:*

$$S_t = \left(S_0 + \frac{\beta}{\sigma}\right) \exp\left[\left((r_d - r_f) - \frac{\sigma^2}{2}\right)t + \sigma W_t\right] - \frac{\beta}{\sigma} \exp\left[(r_d - r_f)t\right] \tag{2}$$

*Proof.* See El-Khatib and Hatemi-J (2017) for a complete proof. We follow the proof of proposition one in that paper, which is based on Ito's lemma.

### 3. The Option Valuation Formula

We consider a European call option on the foreign currency with dynamics following the stochastic deferential equation (1).

**Proposition 2** *Assume that the dynamic of the foreign currency process, $S_T$, is defined by (1), then the premium of an European call option with strike $K$ is provided by the following:*

$$C(S_T, K) = \left(S_0 + \frac{\beta}{\sigma}\right) e^{-r_f T} \Phi(d_1^\beta) - \left(Ke^{-r_d T} + \frac{\beta}{\sigma} e^{-r_f T}\right) \Phi(d_2^\beta), \tag{3}$$

where

$$d_1^\beta = \frac{1}{\sigma\sqrt{T}} \left( \ln\left( \frac{S_0 + \frac{\beta}{\sigma}}{K + \frac{\beta}{\sigma} e^{(r_d - r_f)T}} \right) + \left((r_d - r_f) + \frac{\sigma^2}{2}\right) T \right), \tag{4}$$

and

$$d_2^\beta = \frac{1}{\sigma\sqrt{T}}\left(\ln\left(\frac{S_0 + \frac{\beta}{\sigma}}{K + \frac{\beta}{\sigma}e^{(r_d-r_f)T}}\right) + \left((r_d - r_f) - \frac{\sigma^2}{2}\right)T\right) = d_1^\beta - \sigma\sqrt{T}, \quad (5)$$

where the cumulative probability density function for a standard normal variable is $\Phi(x) = \int_{-\infty}^{x} \frac{e^{-u^2/2}}{\sqrt{2\pi}} du.$

*Proof.* We have

$$C(S_T, K) = E\left[(S_T - K)^+\right]e^{-r_d T} = E\left[(S_T - K)^+\right]e^{-(r_d-r_f)T}e^{-r_f T}. \quad (6)$$

Now, by using Proposition 4 of El-Khatib and Hatemi-J (2017), we have

$$E\left[(S_T - K)^+\right]e^{-(r_d-r_f)T} = \left(S_0 + \frac{\beta}{\sigma}\right)\Phi(d_1^\beta) - \left(Ke^{-(r_d-r_f)T} + \frac{\beta}{\sigma}\right)\Phi(d_2^\beta).$$

Where $d_1^\beta$ and $d_2^\beta$ are given respectively by (4) and (5). We replace the previous expected value in (6) to obtain (3).

The following proposition can be used to determine the premium of a European put option.

**Proposition 3** *Assume that the dynamic of the foreign currency process, $S_T$, is defined by (1), then the premium of an European put option with strike $K$ is given by*

$$P(S_T, K) = \left(S_0 + \frac{\beta}{\sigma}\right)e^{-r_f T}\Phi(d_1^\beta) - \left(Ke^{-r_d T} + \frac{\beta}{\sigma}e^{-r_f T}\right)\Phi(d_2^\beta) + Ke^{-r_d T} - S_0 e^{-r_f T}, \quad (7)$$

where $d_1^\beta$ and $d_2^\beta$ are given by equations (4) and (5) and $\Phi(x) = \int_{-\infty}^{x} \frac{e^{-u^2/2}}{\sqrt{2\pi}} du.$

*Proof.* Similar approach is utilized for the proof as in proposition 2.

## 4. Graphical Presentation of the Predicted Trajectory

In order to compare the performance of our suggested approach with the standard one we present the predicted trajectory for the foreign currency spot rate in Figures (1) and (2). In both graphs we can see that the suggested approach better depicts the effect of increased volatility that characterizes the spot rate in real world compared to the standard approach. This effect is more prominent in Figure 2 in which the parameter value of $\beta$ is higher. These graphs illustrate that the suggested approach accords well with the reality since the exchange rates are more volatile than what the standard approach could capture.

**Figure 1**: The predicted trajectory for the foreign currency spot rate based on standard approach and our suggested approach.

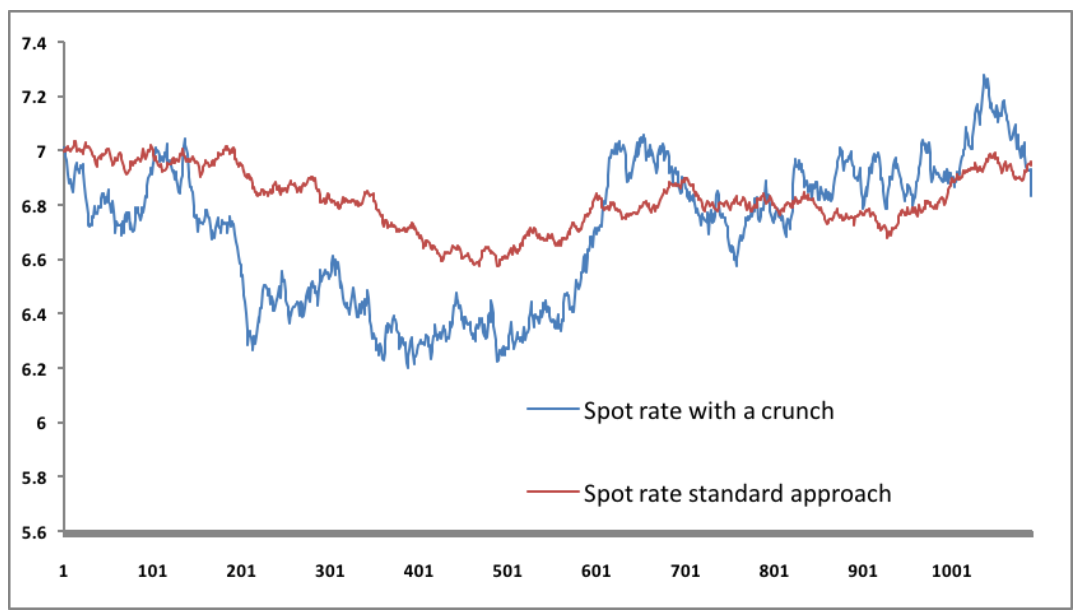

Table 1: Inputs for Figure 1.

| Parameter values | Model Standard | Modified Model |
|---|---|---|
| Spot rate (*S*) | 7 | 7 |
| Volatility (σ) | 0.05 | 0.05 |
| Interest differential ($r_d - r_f$) | 0.02 | 0.02 |
| Time step (Δ*t*) | 0.001 | 0.001 |
| β | 0 | 0.5 |

**Figure 2:** The predicted trajectory for the foreign currency spot rate based on standard approach and our suggested approach. (Higher beta case)

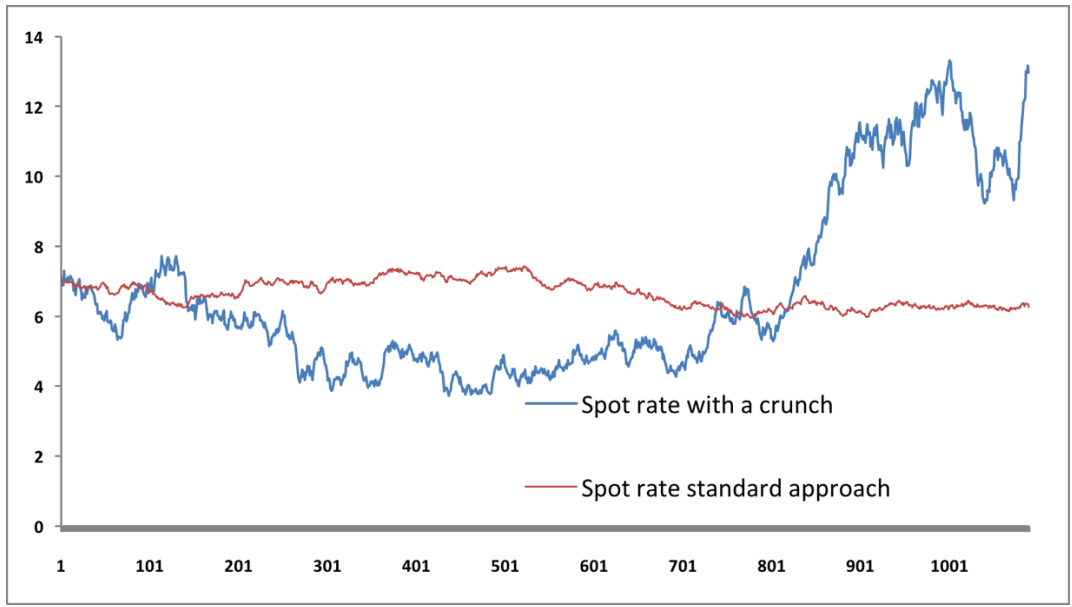

Table 2: Inputs for Figure 2.

| Parameters | Values for Standard Model | Values for Modified Model |
| --- | --- | --- |
| Spot rate ($S$) | 7 | 7 |
| Volatility ($\sigma$) | 0.05 | 0.05 |
| Interest Differential ($r_d - r_f$) | 0.02 | 0.02 |
| Time step ($\Delta t$) | 0.001 | 0.001 |
| $B$ | 0 | 1 |

## 5. An Application

In this section we provide an application in order to demonstrate how the suggested currency option formula performs compared to the standard one. The inputs are presented in Table 3. The estimated premiums for both call and put currency options via each formula are presented in Table 4. As it is evident from these results the alternative formula provides a lower premium for a currency call option and a higher premium for a currency put option compared to the respective premium obtained by the standard formula. These results make sense since during a potential crisis the currency price is expected to go down, which means the likelihood that the call option will be exercised decreases will the opposite is true for the put option in reality.

Table 3. Information on an Original Risky Financial Asset.

| Parameters | Values |
|---|---|
| $\sigma$ | 0.25 |
| $r_f$ | 0.01 |
| $r_d$ | 0.015 |
| $S$ | 2.2 |
| $T$ | 0.75 |
| $K$ | 2.3 |
| $\beta$ | 0.5 |

Notes: Here we have $\sigma$ = the standard deviation, $r_f$ = the foreign risk free rate, $r_d$ = the domestic risk free rate, $S$ = initial currency spot exchange rate, $T$ = time to maturity as a fraction of year, $K$ = exercise currency exchange rate, and $\beta$ = the parameter capturing the effect of the crisis.

Table 4: The Estimation Results of the Premiums for Call and Put Currency Options.

| Estimated values | Standard Model | Modified Model |
|---|---|---|
| Premium for a Call Currency Option | 0.15016 | 0.04940 |
| Premium for a Put Currency Option | 0.24087 | 0.53712 |

## 6. Conclusions

This paper deals with the issue of currency option pricing, which is gaining increasingly importance in globalized financial markets. We suggest a modified version of the standard model for the evaluation of both call and put currency options during a financial crisis. A closed form solution is provided and a mathematic proof for the underlying solution is given. Some simulation results are also generated, which demonstrate the predicted trajectory for the foreign currency spot rate based on our suggested approach accords better with reality compared to the trajectory generated by the standard approach. In addition, an application is provided, which shows that the modified approach results in a lower premium for a call currency option and a higher premium for a put currency option compared to the corresponding values based on the standard approach. This seems to accord well with the reality since in a market that is characterized by a crisis the value of a call option is expected to decrease while the value of a put option is expected to increase assuming the ceteris paribus condition.